\documentclass[12pt,reqno]{article}

\usepackage[authoryear]{natbib}

\usepackage{caption}
\usepackage{subcaption}
\usepackage{bm}
\usepackage{adjustbox,lipsum}
\usepackage{tikz}
\usetikzlibrary{arrows,calc,patterns,math}
\usetikzlibrary{snakes}
\def\centerarc[#1](#2)(#3:#4:#5){ \draw[#1] ($(#2)+({#5*cos(#3)},{#5*sin(#3)})$) arc (#3:#4:#5);}
\definecolor{cornellred}{RGB}{179,27,27} 
\definecolor{cornellblue}{RGB}{55,135,176}
\definecolor{cornellgrey}{RGB}{96,94,92}
\definecolor{cornellgreen}{RGB}{110,180,63}
\definecolor{light-gray}{gray}{0.8}

\usepackage[T1]{fontenc}
\usepackage[latin9]{inputenc}
\usepackage[letterpaper]{geometry}
\usepackage{color}
\usepackage{amsmath}
\usepackage{mathrsfs}
\usepackage{amsthm}
\usepackage{amssymb}
\usepackage{setspace}
\usepackage[plain]{fullpage}           
\usepackage[authoryear]{natbib}
\usepackage[unicode=true,pdfusetitle,
 bookmarks=true,bookmarksnumbered=false,bookmarksopen=false,
 breaklinks=false,pdfborder={0 0 0},pdfborderstyle={},backref=false,colorlinks=true]
 {hyperref}
\hypersetup{
 linkcolor=magenta,citecolor=blue}
 
\usepackage{graphicx}
\graphicspath{ {./graphs/} }
\usepackage{babel}

\makeatother
\usepackage{array}
\usepackage{booktabs}
\usepackage{multirow}
\usepackage{color}

\makeatletter

\usepackage{babel}

\makeatletter

\numberwithin{equation}{section}

\theoremstyle{remark}

\theoremstyle{definition}

\theoremstyle{plain}
\newtheorem{thm}{\protect\theoremname}
\theoremstyle{plain}
\newtheorem{prop}{\protect\propositionname}
\theoremstyle{definition}

\theoremstyle{definition}

\@ifundefined{date}{}{\date{}}
\usepackage{amsmath}
\usepackage{graphicx,psfrag,epsf}
\usepackage{enumerate}
\usepackage{natbib}
\usepackage{url} 

\usepackage{amsthm}
\usepackage{amsfonts}
\usepackage{amstext}

\usepackage{pdflscape}
\usepackage{threeparttable}

\usepackage{natbib}
\makeatother

\providecommand{\conditionname}{Condition}
\providecommand{\definitionname}{Definition}
\providecommand{\examplename}{Example}
\providecommand{\lemmaname}{Lemma}
\providecommand{\propositionname}{Proposition}
\providecommand{\remarkname}{Remark}
\providecommand{\theoremname}{Theorem}

\usepackage{verbatim}

\begin{document}
\sloppy

\title{Statistical Decisions and Partial Identification: \\ With Application to Boundary Discontinuity Design\thanks{To appear in R. Griffith, Y. Gorodnichenko, M. Kandori, and F. Molinari (eds.), \textit{Advances in Economics and Econometrics: Thirteenth World Congress}, Cambridge University Press. Special thanks to Jos\'e Luis Montiel Olea, who collaborated on joint research described here. We also thank Tim Christensen, Han Xu and Kohei Yata for comments, Elliott Serna for outstanding research assistance, and the NSF for financial support under grant SES-2315600.}}

\author{ Chen Qiu\thanks{Department of Economics, Cornell University. Email: cq62@cornell.edu}  \and J\"{o}rg Stoye\thanks{Department of Economics, Cornell University. Email: stoye@cornell.edu}}
\maketitle

\vspace{-.5cm}

\begin{abstract}
We are delighted to respond to the excellent surveys by \citet*{CTYsurvey} and \cite{hirano25}. Our discussion will attempt two things: first, we show how statistical decision theory can be applied to situations with partial identification; second, we connect the surveys' themes by applying these insights to an imagined policy experiment in one of \citeauthor*{CTY25a}'s (\citeyear{CTY25a}) applications.

To do so, we lay out a stylized scenario of statistical decision making under partial identification and, drawing on our own and others' earlier work, provide a complete solution for that scenario. We then apply these results to a hypothetical reduction (modelled on actual policies) in eligibility for educational subsidies. We will see that something of interest can be said, but also that bringing the theory to the application involves some leaps of faith and leaves some questions open. This leads to the final section, where we discuss what we see as the main open challenges in statistical decision theory under partial identification.

\end{abstract}

\newpage{}

\onehalfspacing




\section{A Stylized Decision Problem}\label{sec:main}

Consider a decision maker (DM) who must decide whether to assign a certain treatment to some target population.\footnote{This section follows parts of \citet{stoye2012minimax} and \citet*{MQS}. Readers for whom the language is new might want to consult \citet{hirano25}. For motivations of minimax regret, see in particular \citet{manski2004statistical}, \citet{stoye2012new}, \citet{MQS}, and \citet{hirano25}.} The true effect of this treatment is $\mu^* \in \mathbb{R}$ (status quo welfare being normalized to $0$ throughout), and the DM would ideally assign treatment if $\mu^*>0$, with indifference on the boundary; that is, the \emph{oracle decision rule} is the no-data rule $d^* := \bm{1}\{\mu^*>0\}$.

The DM observes one realization of $\hat{\mu} \sim \mathcal{N}(\mu,\sigma)$, where $\sigma$ is known and $\mu$ is an \textit{observable} or \textit{identified} treatment effect; it is related to $\mu^*$ by $\mu^* \in [\mu-k,\mu+k]$ for some known $k \in [0,\infty)$. Thus, we have a setting of \emph{partial identification}. The substantive motivation might be that the experimental population represents a different location from the treatment population, that it is a population of volunteers, or that the statistical experiment really is a causal inference strategy that may be misspecified or partially identifying to begin with; this last possibility describes our application below. In sum, our model is governed by $(k,\sigma)$, where one may want to think of $k$ as parameterizing the severity of the \emph{identification} problem and $\sigma$ as parameterizing the \emph{estimation} problem. The relative magnitude of $k$ versus $\sigma$ will become accordingly important.

A statistical decision rule is a mapping $d:\mathbb{R} \to [0,1]$ that associates every possible observation $\hat{\mu}$ with a probability of assigning treatment $d(\hat{\mu}) \in [0,1]$.\footnote{Equivalently, one can model an explicit randomization device $U \sim \operatorname{unif}(0,1)$ and then define the treatment rule as $\bm{1}\{U \leq d(\hat{\mu})\}$.} It induces expected welfare of $\mu^* \mathbb{E}[d(\hat{\mu})]$, where the expectation is over the distribution of $\hat{\mu}$ and therefore depends on $(\mu,\sigma)$. We adopt the minimax regret (MMR) criterion, i.e. we evaluate decision rules by their worst-case expected regret
\begin{equation*}\label{eq:MMR}
    \max_{\substack{\mu,\mu^*:\\|\mu^*-\mu|\leq k}}~ \bigl\{~ \underset{\text{oracle welfare}}{\underbrace{\mu^* \bm{1}\{\mu^*>0\}}} - \underset{\text{welfare under }d}{\underbrace{\mu^* \mathbb{E}[d(\hat{\mu})]}} \bigr\} ~= ~   \max_{\substack{\mu,\mu^*:\\|\mu^*-\mu|\leq k}}~ \mu^*\bigl( \bm{1}\{\mu^*>0\}-\mathbb{E}[d(\hat{\mu})]\bigr).
\end{equation*}

Solving such problems can be involved, but this particular example is well understood. The following theorem combines special cases of results in \citet{stoye2012minimax} and \citet{MQS}.

\begin{thm}\label{thm:1}
Under assumptions just stated, the following claims hold.
\begin{itemize}
\item[(i)] If $k\leq \sigma \sqrt{\pi/2} $, then $\bm{1}\{\hat{\mu} >0\}$ is MMR optimal, uniquely so (up to a.s. agreement in sample space) if the inequality is strict.
\item[(ii)] Else, MMR is attained by
\begin{equation*}\label{eq:linear.rule}
d^*_{\text{linear}}(\hat{\mu}):=\begin{cases}
0, & \hat{\mu}<-k^*,\\
\frac{\hat{\mu}+k^*}{2k^*}, & -k^*\leq\hat{\mu}\leq k^*,\\
1, & \hat{\mu}>k^*,
\end{cases}
\end{equation*}
where $k^{*}<k$ is the unique strictly positive  solution of
\begin{eqnarray*}
\frac{k^*}{2k} - \frac{1}{2} + \Phi\left(-\frac{k^*}{\sigma}\right) =0.
\label{eq:rho.star.main.proof} 
\end{eqnarray*} 
\item[(iii)] In case (ii), any decision rule that attains MMR and that coincides (as function of $\hat{\mu}$) with the c.d.f. of a symmetric (about $0$), unimodal distribution on $\mathbb{R}$ must randomize on an interval containing $[-k^*,k^*]$.
\end{itemize}
\end{thm}

\begin{figure}[t]
		\begin{subfigure}[b]{.5\linewidth}
			\begin{adjustbox}{max width=\textwidth}
\tikzstyle{line} = [draw, -latex']
\begin{tikzpicture}[node distance = 2cm, auto]
\linespread{0.9}
\draw[->] (0,0) -- (0,3.5) node[left] {$a$};
\draw[-] (1,0) -- (0,0) node[below] {$0$};
\draw[-] (0,3) -- (0,3) node[left]{$1$};
\draw[->] (-4,0) -- (4,0) node[below] {$\hat{\mu}$};
\draw[dashed] (-4,3) -- (4,3) ;
\draw[-] (0,3) -- (0,3) node[left]{$1$};
\draw [fill] (-.7,0) circle [radius=2pt] node[below] {$-k$};
\draw [fill] (.7,0) circle [radius=2pt] node[below] {$k$};
\draw[cornellred,very thick] (-4,0) -- (0,0) -- (0,3) -- (4,3);
\end{tikzpicture}
\end{adjustbox}
		\end{subfigure}
		\begin{subfigure}[b]{.5\linewidth}
		\begin{adjustbox}{max width=\textwidth}
\begin{tikzpicture}[node distance = 2cm, auto]
\linespread{0.9}
\draw[->] (0,0) -- (0,3.5) node[left] {$a$};
\draw[-] (1,0) -- (0,0) node[below] {$0$};
\draw[-] (0,3) -- (0,3) node[left]{$1$};
\draw[->] (-4,0) -- (4,0) node[below] {$\hat{\mu}$};
\draw[dashed] (-4,3) -- (4,3) ;
\draw[-] (0,3) -- (0,3) node[left]{$1$};
\draw [fill] (-3,0) circle [radius=2pt] node[below] {$-k$};
\draw [fill] (-2,0) circle [radius=2pt] node[below] {$-k^*$};
\draw [fill] (2,0) circle [radius=2pt] node[below] {$k^*$};
\draw [fill] (3,0) circle [radius=2pt] node[below] {$k$};
\draw[dashed] (2,0) -- (2,3) ;
\draw[cornellred,very thick]  (-4,0) --(-2,0) -- (2,3) -- (4,3);
\end{tikzpicture}
\end{adjustbox}
\end{subfigure}
\caption{MMR treatment choice as function of $\hat{\mu}$; see Theorem \ref{thm:1}. Left: For small $k$ (or large $\sigma)$, model uncertainty is ignored. Right: For large $k$, there is randomization, but the recommended decision still effectively ignores model uncertainty for some signals $\hat{\mu}$ (with absolute value between $k^*$ and $k$) s.t. estimated bounds on the treatment effect do not identify its sign.}
\label{fig:T1}
\end{figure}
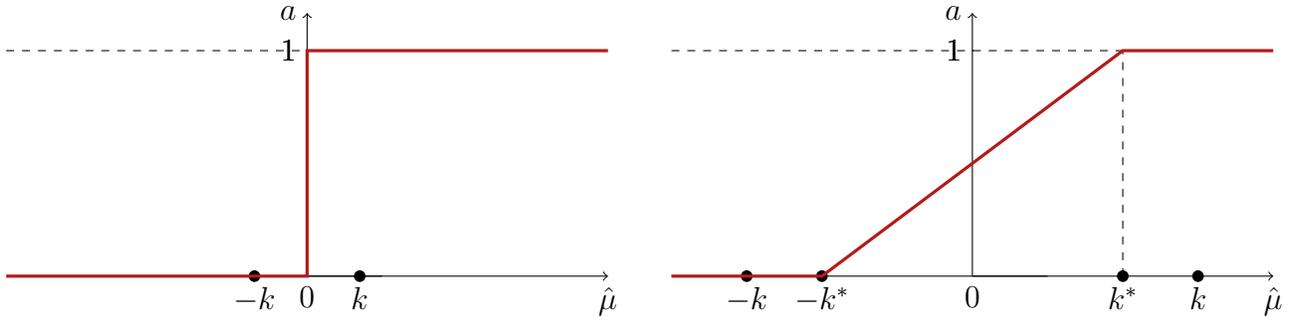
The primary way in which (i) and (ii) were established is through guessing and verifying Nash equilibria of a zero-sum game against a malicious Nature that can randomize over true parameter values. Of note, if such an equilibrium is found, then being an equilibrium strategy is also \textit{necessary} for a decision rule to attain MMR; this is used in (iii).  

Some interesting aspects of Theorem \ref{thm:1} are visualized in Figure \ref{fig:T1}. First, in the point-identified edge case where $k=0$, it is expected (and easily shown) that $\bm{1}\{\hat{\mu} >0\}$ is the unique MMR solution. Much less obviously, this is also true for sufficiently small $k>0$. That is, \emph{it is optimal to completely ignore small but positive model uncertainty}. Second, while the MMR rule from part (ii) is not unique, it further justifies ignoring model uncertainty in some other cases, namely if the estimated bounds $[\hat{\mu}-k,\hat{\mu}+k]$ either do not contain zero (thus, the estimator implies a unique sign of $\mu^*$, though note that we are silent on significance) or contain zero but are sufficiently lopsided toward one side of it (thus, the estimator is merely suggestive of the sign of $\mu^*$). Finally, within a large regularity class of decision rules that attain MMR, $d^*_{linear}$ maximizes the region in sample space on which model uncertainty is ignored; in particular, it is not possible to attain MMR without randomization. Indeed, a maybe controversial feature of MMR is that it may prescribe randomization (which may be population-wide all-or-nothing or fractional, i.e. ex post unequal) well beyond ``small'' decision frontiers; for a defense of this feature, see especially \citet{Manski09}.

\section{Application to Boundary Discontinuity Design}\label{sec:app}

Consider \citeauthor{CTY25a}'s (\citeyear{CTY25a}, building on \citet*{Colombia}) application to educational subsidies. The bivariate running variable is $\mathbf{X}:=(X_{1},X_{2})$, where $X_{1}$ is a student's ability measure and $X_{2}$ is family wealth. After recentering, the treatment assignment boundary is 
\[
\mathscr{B}=\left\{ (x_{1},x_{2})\in\mathbb{R}^{2}:\left(x_{1}\geq0,x_{2}=0\right)\text{ or }\left(x_{1}=0,x_{2}\geq0\right)\right\}.
\]
A student is assigned to treatment if both measures are above the cutoff. The outcome of interest $Y\in\{1,0\}$ indicates whether the student attended college after treatment assignment. 

\cite{CTY25a} study how to estimate treatment effects for the subpopulation along $\mathscr{B}$. Here, we ask a slightly different question: should a policy maker change the eligibility of the program to maximize welfare? As a concrete illustration, suppose the policy maker contemplates increasing the ability threshold ($X_{1}$) of the program by a fixed number $\Delta>0$. These increases actually occurred in Colombia subsequently to \citeauthor{Colombia}'s (\citeyear{Colombia}) data collection, and the values of $\Delta$ used below mimic actual policies. In one-dimensional regression discontinuity design, analyzing them would correspond to \citeauthor{yata2021}'s (\citeyear{yata2021}) main empirical application. 

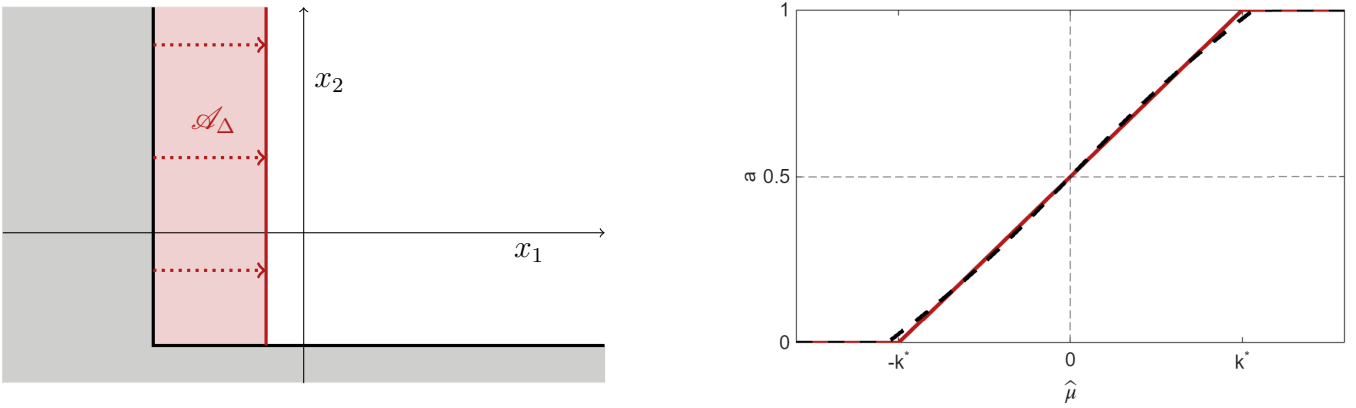
\begin{figure}[t]
\begin{subfigure}[b]{.55\linewidth}
		\begin{adjustbox}{max width=\textwidth}
\begin{tikzpicture}[node distance = 2cm, auto]
\linespread{0.9}
\filldraw[cornellgrey!30] (-4,-2) -- (-4,3) -- (-2,3)  -- (-2,-1.5) --  (4,-1.5)-- (4,-2) -- (-4,-2);
\filldraw[white] (-4,-2) -- (4,-2) -- (4,-2.3)  -- (-4,-2.3) --   (-4,-2);
\filldraw[cornellred!20]  (-2,-1.5) --  (-2,3)-- (-.5,3) -- (-.5,-1.5);
\draw[->] (-4,0) -- (3,0) node[below] {$x_1$} -- (4,0);
\draw[->] (0,-2) -- (0,2) node[right] {$x_2$} -- (0,3);
\draw[very thick] (-2,3) -- (-2,-1.5) -- (4,-1.5);
\draw[very thick,cornellred] (-.5,3) -- (-.5,-1.5) ;
\draw[->,very thick,dotted,cornellred] (-2,2.5) -- (-.5,2.5);
\draw[->,very thick,dotted,cornellred] (-2,1) -- (-.5,1);
\draw[->,very thick,dotted,cornellred] (-2,-.5) -- (-.5,-.5);
\draw[cornellred] (-1.2,1.5) node{$\mathscr{A}_\Delta$} ;
\end{tikzpicture}
\end{adjustbox}
\end{subfigure}
\begin{subfigure}[b]{.45\linewidth}
			\begin{adjustbox}{max width=\textwidth}
    \includegraphics[trim=1.8cm 7.8cm 2.5cm 8cm,clip=true,scale=.6]{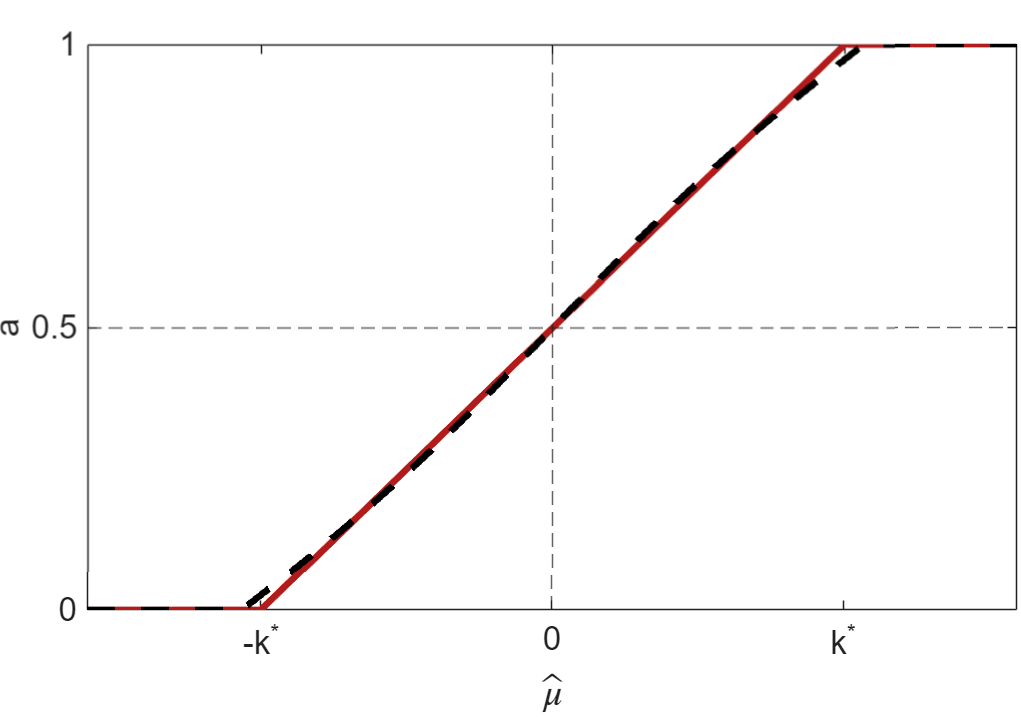}
\end{adjustbox}
		\end{subfigure}
\caption{Left: A policy experiment in the setting of \citet{CTY25a}. Eligibility for treatment follows a double threshold rule; we consider increasing the first threshold. Right: Approximated (black, dashed) and exact (red, solid; compare the right panel of Figure \ref{fig:T1}) MMR rule from Theorem \ref{thm:1}. See \citet[][Figure 1]{fernandez2024epsilon} for parameter values.}
\label{fig:RDD}
\end{figure}

The policy maker's action $a\in[0,1]$ is the probability of increasing the ability threshold by $\Delta$. Let $\tau(\mathbf{x}):=\mathbb{E}[Y(1)-Y(0)|\mathbf{X}=\mathbf{x}]$
be the conditional average treatment effect and $\mathscr{A}_\Delta:=\left\{ (x_{1},x_{2})\in\mathbb{R}^{2}:x_{1}\in[0,\Delta),x_{2}\geq 0\right\}$ 
the policy relevant subpopulation (see the left panel of Figure \ref{fig:RDD}), $c$ and $b$ the average cost and benefit of the program per treatment assignment (measured in monetary terms), respectively. Then the policy relevant payoff parameter is
\begin{equation*}
\mu^{*}=c-b\frac{\int_{\mathbf{x}\in\mathscr{A}_\Delta}\tau(\mathbf{x})dF(\mathbf{x})}{\int_{\mathbf{x}\in\mathscr{A}_\Delta}dF(\mathbf{x})},
\end{equation*}
where $F(\cdotp)$ denotes the cdf of $\mathbf{X}$. We assume $c$ and $b$ are known; in the application, they are taken from \citeauthor{Colombia}'s (\citeyear{Colombia}, Appendix 8) cost-benefit analyses.  

Boundary regression discontinuity design allows us to identify treatment effects on $\mathscr{B}$ but not on $\mathscr{A}_\Delta$. Therefore, we adopt a partial identification approach. Following \citet[][see also \citet*{christensen2022optimal}; \cite{ishihara2021,MQS}]{yata2021}, we impose a Lipschitz constraint on $\tau(\cdot)$:
\begin{equation}\label{eq:lipschitz}
\left|\tau(\mathbf{x})-\tau(\mathbf{y})\right|\leq C\left\Vert \mathbf{x}-\mathbf{y}\right\Vert ,\forall\mathbf{x},\mathbf{y}
\end{equation}
for some known constant $C\geq 0$,  where $\left\Vert \cdotp\right\Vert $ is the Euclidean norm.\footnote{For brevity, we make choices that allow us to directly invoke \cite{CTY25a}. Similar analyses would apply after (arguably more natural) separate imposition of Lipschitz constraints on $\mathbb{E}[Y_j|\bm{X}=\bm{x}],j=0,1$. Also, as with matching estimators, the choice of norm is not obvious; see especially \citet[][Section 18.5]{imbens2015causal}.}
Then $\mu^*$ is tightly bounded above and below by the values of
\begin{eqnarray}\label{eq:tight_bounds}  
 \sup_{\tau(\cdot)} /\inf_{\tau(\cdot)}  \left\{c-b\frac{\int_{\mathbf{x}\in\mathscr{A}_\Delta}\tau(\mathbf{x})dF(\mathbf{x})}{\int_{\mathbf{x}\in\mathscr{A}_\Delta}dF(\mathbf{x})}\right\}
\text{    s.t.  }\left|\tau(\mathbf{x})\right|\leq 1\text{   and   \eqref{eq:lipschitz} holds for all }\mathbf{x},\mathbf{y} \in \mathscr{A}_\Delta \cup \mathscr{B}. 
\end{eqnarray}
One could in principle evaluate decision rules based on these bounds, but executing this would be beyond this note's scope. However, consider weakening the constraints to
\begin{equation}\label{eq:direction.Lip}
|\tau(\mathbf{x})-\tau(\mathbf{y})|\leq C|x_1-y_1| \text{ for all } \mathbf{x},\mathbf{y}\in \mathscr{A}_\Delta \cup \mathscr{B} \text{ s.t. } x_2=y_2,
\end{equation}
a Lipschitz constraint only in the direction of the first coordinate for units sharing the same value of the second.  Then $\tau(\mathbf{x})$ for $\mathbf{x}=(x_1,x_2)\in \mathscr{A}_\Delta$ is directly constrained only by $|\tau(\mathbf{x})-\tau(\mathrm{P}\mathbf{x})|\leq C|x_1-0|=\Vert\mathbf{x}-\mathrm{P}\mathbf{x}\Vert$, where $\mathrm{P}\mathbf{x}:=(0,x_{2})$. Thus, we have:

\begin{prop}\label{prop:1}
If $\tau(\cdot)$ is such that \eqref{eq:direction.Lip} holds, then $\mu^{*}\in[\mu-k,\mu+k]$, with
\begin{align}\label{eq:prop1}
\mu  :=c-b\frac{\int_{\mathbf{x}\in\mathscr{A}_\Delta}\tau(\mathrm{P}\mathbf{x})dF(\mathbf{x})}{\int_{\mathbf{x}\in\mathscr{A}_\Delta}dF(\mathbf{x})},\quad
k :=bC\frac{\int_{\mathbf{x}\in\mathscr{A}_\Delta}\left\Vert \mathbf{x}-\mathrm{P}\mathbf{x}\right\Vert dF(\mathbf{x})}{\int_{\mathbf{x}\in\mathscr{A}_\Delta}dF(\mathbf{x})}.
\end{align}
\end{prop}
The expressions in Proposition \ref{prop:1} resemble Theorem \ref{thm:1}. Note that $\frac{\int_{\mathbf{x}\in\mathscr{A}_\Delta}\tau(\mathrm{P}\mathbf{x})dF(\mathbf{x})}{\int_{\mathbf{x}\in\mathscr{A}_\Delta}dF(\mathbf{x})}$ is what \cite{CTY25a} call a boundary average treatment effect (BATE) and is point identified. We next make two leaps of faith. First, suppose that $F(\cdot)$ is known. For example, our sample might be the population of interest or the distribution of $\mathbf{X}$ might be estimated on a much larger sample, so that we are willing to abstract from the corresponding sample uncertainty. In this case, both $k$ and the implied weights on the BATE are known, although the latter are not the ones from \citeauthor{CTY25a}'s (\citeyear{CTY25a}) empirical application. Second, we assume that Theorems 3 and 4 in \cite{CTY25a} \textit{exactly} describe the statistical behavior of $\hat{\mu}$, so that we can take $\hat{\mu}\sim N(\mu,V)$, where $V$ is furthermore known. Theorem \ref{thm:1} then applies and can give concrete guidance on treatment choice. For example, MMR treatment choice is not randomized if $bC\int_{\mathscr{A}_\Delta}\left\Vert \mathbf{x}-\mathrm{P}\mathbf{x}\right\Vert dF(\mathbf{x})\leq \sqrt{V\pi/2}\int_{\mathscr{A}_\Delta}dF(\mathbf{x})$. In addition, in this case, treatment assignment is guided by the estimated sign of $\mu$, the estimand corresponding to imposing \eqref{eq:direction.Lip} with $C=0$, a point identifying assumption. In other words, for $C$ small enough or $V$ large enough, the MMR rule performs as if model uncertainty had been assumed away. In Table \ref{table:cattaneo}, we report $\hat{\mu},V,k$ and the MMR optimal policy using data from \cite{CTY25a} for a range of values of $C$ and $\Delta$. Here, the choice of $C\in\{.005,.01,.015\}$ is admittedly ad hoc, but even the smallest value exceeds a guesstimate of $C$ obtained from inspecting estimators along the horizontal segment of $\mathscr{B}$;    $\Delta$ is calibrated to actual subsequent policy changes in Colombia as reported by \citet{Colombia}; $c=\$12234$ and $b=\$59744$ follow \citeauthor{Colombia} (\citeyear{Colombia} Appendix 8). We also note that the constraint $|\tau(\bm{x})|\leq1$, which we dropped for simplicity, is not close to binding in the application.

\begin{table}[http]
\begin{centering}
\begin{tabular}{c|ccc|ccc|ccc}
\toprule
 & \multicolumn{3}{c|}{$\Delta=8$}
 & \multicolumn{3}{c|}{$\Delta=32$}
 & \multicolumn{3}{c}{$\Delta=38$} \\
$C$
& $0.005$ & $0.01$ & $0.015$
& $0.005$ & $0.15$ & $0.015$
& $0.005$ & $0.01$ & $0.015$ \\

$\hat{\mu}$
&  & $-6338.6$ & 
&  & $-6356.7$ & 
&  & $-6358.8$ &  \\
$V$
&  & $1241210$ & 
&  & $1243899$ & 
&  & $1243966$ &  \\
$k$
& $973.7$ & $1947.4$ & $2921.1$
& $3603.7$ & $7207.5$ & $10811.2$
& $4079.0$ & $8158.1$ & $12237.1$ \\
$d_{linear}^{*}(\hat{\mu})$
& $0$ & $0$ & $0$
& $0$ & $0.0590$ & $0.2060$
& $0$ & $0.1102$ & $0.2402$ \\
\bottomrule
\end{tabular}
\par\end{centering}
\caption{Values of $(\hat{\mu},V,k)$ using data from \cite{CTY25a} and \citet{Colombia}.}\label{table:cattaneo}
\end{table}
To be clear, this result allows two interpretations: If we truly impose Lipschitz continuity only in the fashion of \eqref{eq:direction.Lip} (and subject to the aforementioned leaps of faith), then the result is exact. Alternatively, since we effectively exploited \eqref{eq:lipschitz} only in one coordinate, one can interpret \eqref{eq:prop1} as outer approximation of the bounds in \eqref{eq:lipschitz}. Because this corresponds to relaxing the adversary's constraints in the statistical game, the decision rule derived from Theorem \ref{thm:1} is still guaranteed to minimize a (hopefully informative) upper bound on worst-case expected regret. Resorting to bounds on regret is a common strategy for intractable decision problems; compare \citet[][Section 3.3]{manski2004statistical} and recently \citet*{kmedian}. Under either interpretation, the tendency is for MMR to caution against increasing eligibility thresholds; in fact, for most parameter values the decision recommendation is a nonrandomizing $0$.

We close this section by mentioning possible extensions. It would be of obvious interest to develop the tight bounds in \eqref{eq:tight_bounds} as well as refined or different bounds using different assumptions. For example, \citeauthor{deaner2025extrapolationregressiondiscontinuitydesign}'s (\citeyear{deaner2025extrapolationregressiondiscontinuitydesign}) assumption of comononicity between mean potential outcomes as functions of covariates would by itself induce non-nested partial identification (unless we observe the entire range of both functions) and could be combined with smoothness conditions. We note that, in all of these cases, even asymptotic normality of estimators would not be obvious. Next, rather than evaluating one particular threshold shift, one could investigate optimal shifts; for example, see \citet{Crippa25} in a point identified and \citet{Manski25medical} in a partially identified setting. Finally, our development relied on several leaps of faith, some of which we address next.

\section{What's Next?}\label{sec:whatsnext}

Finding minimax decision rules is challenging. In an important contribution, \citet{yata2021} generalizes Theorem \ref{thm:1} to a centrosymmetric parameter space and vector-valued signal. But this is still restrictive; furthermore, while Yata provides essential dimension reduction through analytic arguments, a final step has to be resolved either numerically (as in Yata's own application) or by guessing and verifying the equilibrium of an, albeit simplified, statistical game (as in \citet[][Proposition 1]{MQS}). \citet{ishihara2021} propose a simplified class of decision rules and then also combine analytical dimension reduction with numerical steps. Other examples of ``mostly closed-form'' derivations of MMR decision rules with partial identification include \citet*{twopointprior}, \citet*{kitagawa2023treatment}, \citet{Manski07}, and \citet{Stoye07}; the last two abstract from sampling uncertainty. The list is short, and inspection of the papers further reveals that the ``guess and verify'' approach must encounter limitations of scale.

It therefore appears essential to discover optimal decision rules at scale. Two salient approaches (that can, of course, be combined) are: (i) to simplify the game through asymptotic approximation and (ii) to automate discovery of minimax rules in tractable statistical decision problems.

An obvious starting point for (ii) is the characterization of minimax decision rules as equilibrium strategies in statistical games. Can we leverage recent advances in finding Nash equilibria of zero-sum games? This question was recently and independently pursued by \cite*{fernandez2024epsilon} and \citet{guggenberger2025numerical}.

\citet{guggenberger2025numerical} assume that both the parameter space and the space of decision rules are finite; each may in practice reflect discretization of infinite spaces. The statistical game then becomes finite. \citet{guggenberger2025numerical} approximately solve it by a close variant of fictitious play and also clarify conditions under which discretization can be made ``dense.''

\cite{fernandez2024epsilon} discretize only the strategy space. Building on recent advances in optimization theory \citep*{arora2012multiplicative}, they give guaranteed bounds on true minimax loss as well as rate guarantees that are furthermore essentially optimal for first-order (i.e., using only up to first derivatives) algorithms. However, in practice they require an oracle that rapidly computes (at least approximately) nature's best response in the statistical game. 

Theorem \ref{thm:1} corresponds to the first application in \cite{fernandez2024epsilon}; the resulting approximation is visualized in the right panel of Figure \ref{fig:RDD}. Attention was restricted to decision rules that increase in $\hat{\mu}$, a class that includes a true solution and that is easily approximated as convex hull of a fine grid of threshold rules. The approximation is evidently successful, as is also reflected in tight bounds (not reported here) on true minimax regret loss. In sum, the approach works well in this example. We have no doubt that the same would be true for \citet{guggenberger2025numerical}. Practical feasibility of these approaches in more difficult cases remains to be investigated.   

Regarding strategy (i), Theorem \ref{thm:1} nominally provides a finite sample solution to a statistical game, but in reality relies on a large sample approximation through assuming normal errors. Such hybrid constructions are common in related work including our own \citep{ishihara2021,kitagawa2023treatment,MQS,stoye2012minimax,tetenov2012statistical,yata2021}. However, it would be ideal to embed them in a rigorous asymptotic framework. In a treatment choice context, pioneering contributions are due to \citet{HiranoPorter2009,HiranoPorter2020}, who develop local asymptotic theory for treatment assignment rules using \citeauthor{le2012asymptotic}'s (\citeyear{le2012asymptotic}) limits of experiments framework, but they assume point identification.  
Recent efforts that aim to extend the theory of \citet{HiranoPorter2009} to partially identified settings include \cite{christensen2020robust,christensen2022optimal,kido2023locally,xu25asymptotic}. Focusing on non-randomized decision rules, \cite{christensen2020robust,christensen2022optimal} profile out the partially  identified parameter in the loss function and consider a conditional $\Gamma$-minimax criterion. \cite{xu25asymptotic} also considers profiled loss functions  but focuses on Bayes and minimax optimality analyses. \cite{kido2023locally} profiles out the partially identified parameter in the risk function; see also \cite{song2014point}. \cite{yata2021} has a global asymptotic argument for his derived optimal rules following the asymptotic framework considered by \cite{armstrong2018optimal}. This literature is still in flux; we suspect this illustrates the delicacy of combining partial identification with asymptotic experiments, with different, plausible ways of conceptualizing the latter informing different notions of asymptotic optimality as well as optimal decision rules.

\bibliographystyle{ecta}
\bibliography{Vol1_ch3_bib}

\end{document}